\documentstyle[prl,twocolumn,aps]{revtex}

\def\beq{\begin{equation}}
\def\eeq{\end{equation}}
\def\6{\langle}
\def\9{\rangle}

\def\bk{{\bf k}}
\def\bq{{\bf q}}
\def\bv{{\hat{\bf v}}}
\def\bw{{\hat{\bf w}}}
\def\hbk{{\hat\bk}}
\def\hbq{{\hat\bq}}
\def\sR{{\cal R}}
\def\qk{_{\bq\bk}}

\begin{document}

\title{Wigner's little group and Berry's phase for massless particles}
\author{Netanel H. Lindner and Asher Peres}
\address{Department of Physics, Technion---Israel Institute of
Technology, 32000 Haifa, Israel}
\author{Daniel R. Terno}
\address{Perimeter Institute for Theoretical Physics, Waterloo,
Ontario, Canada N2J 2W9}

\maketitle
\begin{abstract}
The ``little group'' for massless particles (namely, the Lorentz
transformations $\Lambda$ that leave a null vector invariant) is
isomorphic to the Euclidean group E2: translations and rotations
in a plane. We show how to obtain explicitly the rotation angle of
E2 as a function of $\Lambda$ and we relate that angle to Berry's
topological phase. Some particles admit both signs of helicity,
and it is then possible to define a reduced density matrix for their
polarization. However, that density matrix is physically meaningless,
because it has no transformation law under the Lorentz group, even
under ordinary rotations.

\bigskip

\end{abstract}
\pacs{03.65.Ta, 03.30.+p.}
%\maketitle

Eugene Wigner considered his paper ``On unitary representations of
the inhomogeous Lorentz group'' \cite{wigner} as his most
important contribution to physics \cite{priv}. The key feature in
that article was the introduction of a {\it little group\/},
namely a subgroup under which a standard vector $s_\mu$ is
invariant. For example, a timelike $s_\mu$ is (1,0,0,0) and the
little group is the familiar rotation group SO(3). A null standard
vector can be taken as (1,0,0,1). Spacelike standard vectors have
no physical interest.

To find explicitly the little group that corresponds to the null
$s_\mu$, let us introduce an auxiliary complex null vector $m_\mu$
such that \cite{aop}

\beq m_\mu\,s^\mu=m_\mu\,m^\mu=0, \eeq \beq m_\mu^*\,m^\mu=-1,
\eeq and a real null vector $n_\mu$ that satisfies \beq
n_\mu\,m^\mu=0,\eeq \beq n_\mu\,s^\mu=1.\eeq All these properties
are manifestly Lorentz invariant. Moreover they still hold under
the transformation

\beq m_\mu\to e^{i\xi}\,m_\mu+\gamma s_\mu, \eeq \beq n_\mu\to
n_\mu+e^{i\xi}\gamma^* m_\mu+
      e^{-i\xi}\gamma\,m^*_\mu+|\gamma|^2 s_\mu,\eeq
where $\gamma=\alpha+i\beta$ is any complex number.  Therefore the
above transformation of $m_\mu$ and $n_\mu$ is a subgroup of the
Lorentz group. We thereby obtain a representation of the generic
little group element

\beq g=S(\alpha,\beta)R_z(\xi),\label{decl} \eeq
where the $S(\alpha,\beta)$ form a subgroup which is isomorphic to the
translations of a Euclidean plane and $R_z(\xi)$ is a rotation around
the origin of that plane, which in this case is also a rotation around
the $z$-axis. We see that the little group is E2, as shown in various
ways by other authors \cite{halpern,bog1,bog2,weinberg}.

In the case of infinitesimal transformations in Minkowski space,
we have \beq g={\bf1}+\alpha A+\beta B+\xi M_{12},
\label{infi}\eeq where $M_{12}\equiv J_3$, and $A$ and $B$ are
commuting generators of translations in the 12-plane, for example
\cite{ham},

\beq A=M_{01}+M_{31}\qquad \mbox{and} \qquad B=M_{02}+M_{32}.
\label{AB} \eeq

In general, let $\Lambda^\mu_\nu$ denote the Lorentz
transformation matrix and $k^\nu$ an arbitrary momentum.  The task
is to find the little group matrix $W^\mu_\nu$ that corresponds to
$\Lambda^\mu_\nu$ and $k^\nu$, namely

\beq W(\Lambda,k)=L^{-1}(\Lambda k)\Lambda L(k), \label{wdef} \eeq
where spacetime indices were omitted for brevity, and $L(k)$ is
the standard Lorentz transformation (next equation) that converts
$s$ to an arbitrary momentum $k$.  For example, in the case of
massive particles for which the little group is SO(3), if
$\Lambda$ is an ordinary rotation then $W$ is the same rotation
irrespective of $k$. However if $\Lambda$ is a boost, then $W$
does depend on $k$. Likewise the various terms in (\ref{decl}) may
depend on $k$.

In this Letter, we shall examine the group properties of massless
particles.  For $s=(1,0,0,1)$, the {\it standard Lorentz
transformation\/}, as defined in \cite{weinberg}, is

\beq L(k)=R(\hbk)B_z(|\bk|),\label{standtf}\eeq where $B_z(|\bk|)$
is a boost along the $z$-axis with velocity $(1-\bk^2)/(1+\bk^2)$,
and $R(\hbk)$ is the standard rotation that carries the $z$-axis
into the direction of the unit vector $\hbk$. Again following
\cite{weinberg}, if the direction of $\hbk$ is given by spherical
angles $\theta$ and $\phi$, the {\it standard rotation\/}
$R(\hbk)$ consists of a rotation $\theta$ around the $y$-axis,
followed by a rotation $\phi$ around the $z$-axis:

\beq R(\hbk)=R_z(\phi)R_y(\theta). \label{standrot}\eeq
In these formulas, all Lorentz transformations are passive, even
if we occasionally use an active wording.

We now prove by a {\it classical geometric argument\/} that, if
the Lorentz transformation is a pure rotation ($\Lambda=\sR$),
then it follows from Eq.~(\ref{wdef}) that $S(\alpha,\beta)$
simply is the unit matrix, and therefore $\alpha=\beta=0$. We also
give a simple expression for the rotation angle $\xi$. From the
definition of the little group element we have

\beq
W(\sR,\bk)=B_z^{-1}(|\bk|)R^{-1}(\sR\hbk)\sR R(\hbk)B_z(|\bk|).
\eeq
Since the action of $R^{-1}(\sR\hbk)\sR R(\hbk)$ leaves the
$z$-axis invariant, it is equivalent to some rotation
$R_z(\varpi)$ around that axis,

\beq W(\sR,\bk)=B_z^{-1}(|\bk|)R_z(\varpi)B_z(|\bk|)=R_z(\varpi),
\label{newdecom} \eeq
and since $W(\sR,k)$ is a special case of (\ref{wdef}), it follows
that in Eq.~(\ref{decl}) we have $S^\mu_\nu=\delta^\mu_\nu$ and
$\alpha=\beta=0$. Equation~(\ref{decl}) also gives $\xi=\varpi$,
and the remaining problem is to find the explicit value of this angle.

To clarify the origin of the phase $\xi$ in Eq.~(\ref{decl}), we
note that any rotation in a three-dimensional space can be described
by two angles that give the direction of the rotation axis, and a
third angle that gives the amount of rotation around that axis. A
rotation from $\bk$ to $\bq=\sR\bk$ can be performed in many ways
(denoted below by $R\qk$), in addition to the given $\sR$ that we are
seeking to decompose. Since all such rotations satisfy

\beq \sR\bk=R\qk\bk, \qquad \sR^{-1}\bq=R^{-1}\qk\bq, \eeq
the difference between them is a rotation that preserves $\hbq$,
if done after $R\qk$, or a rotation that preserves $\hbk$, if done
before $R\qk$.  In particular, $\bq=\sR R^{-1}\qk\bq$, so that

\beq \sR R^{-1}\qk=R_\hbq(\omega),\label{defomega} \eeq
where $R_\hbq(\omega)$ is a rotation around $\hbq$. Among the
infinity of possible $R\qk$ we choose

\beq R\qk=R(\hbq)R^{-1}(\hbk),\eeq
where $R(\hbq)$ and $R(\hbk)$ are standard rotations, as in
Eq.~(\ref{standrot}). It follows that

\beq \sR=R_{\sR\hbk}(\omega)R(\sR\hbk)R^{-1}(\hbk), \label{defsR} \eeq
where $R_{\sR\hbk}(\omega)$ is a rotation around $\sR\hbk$, while
$R(\sR\hbk)$ and $R(\hbk)$ are the standard rotations that carry
the $z$-axis to $\sR\hbk$ and $\hbk$, respectively. We can thus
consider Eq.~(\ref{defomega}) as the definition of
$R_{\sR\hbk}(\omega)$.

Substituting this decomposition into Eq.~(\ref{newdecom}), we
obtain

\beq W(\sR,k)=R^{-1}(\sR\hbk)R_{\sR\hbk}(\omega)R(\sR\hbk)=R_z(\varpi),
\label{rdeco}\eeq
and we conclude that $\xi=\varpi=\omega$.

To obtain the rotation angle under a general Lorentz transformation,
we decompose the latter into two rotations and a standard boost $B_z$
along the $z$-axis \cite{tung}:

\beq \Lambda=\sR_2B_z(u)\sR_1. \eeq
As shown below, $B_z$ alone does not lead to a phase rotation.
Therefore,

\beq \xi=\omega_1+\omega_2, \eeq
where both $\omega_1$ and $\omega_2$ are due to the rotations and are
given by Eq.~(\ref{defsR}). 

We now prove that $B_z$ alone induces no phase rotation \cite{adami}. 
Consider a pure boost along the $z$-axis, $\Lambda=B_z(u)$, and a
generic null vector $k=(|\bk|,\bk)$, where

\beq \bk = |\bk|(\sin\theta\sin\phi, \sin\theta\cos\phi, \cos\theta).
\eeq
We define $q=B_z(u)k=(|\bq|,\bq)$ where

\beq \bq = |\bq|(\sin\theta'\sin\phi,
\sin\theta'\cos\phi,\cos\theta').  \eeq
Note that the angle $\phi$ is the same for $\bk$ and for $\bq$. Thus

\beq R^{-1}(\hbq)=R^{-1}_y(\theta')R_z^{-1}(\phi). \label{rotq} \eeq
We now substitute Eqs.~(\ref{standrot}) and (\ref{rotq}) into
Eq.~(\ref{wdef}), to obtain

\beq
W(B_z(u),k)=B_z^{-1}(|\bq|)R_y^{-1}(\theta')B_z(u)R_y(\theta)B_z(|\bk|)
\label{wBz}, \eeq
where we used $R_z^{-1}(\phi)B_z(u)R_z(\phi)=B_z(u)$.

Consider now the effect of the little group element (\ref{wBz}) on  the
spacelike vector $y=(0,0,1,0)$. That vector is not affected by a boost
in the $z$ direction, nor by a rotation around the $y$-axis. Therefore

\beq W(B_z(u),k)y=y,  \eeq
so that in this case $\xi$ is either $0$ or $2\pi$. Since for $u=0$
we expect $\xi=0$, by continuity $\xi=0$ for all $u$.

Note that although $B_z(u)$ alone does not lead to a phase rotation,
it can affect the value of $\omega_2$, since it indirectly appears in
the definition of $\sR_2$. Indeed, if we decompose $\sR_2$  as in
Eq.~(\ref{defsR}), we obtain

\beq \sR_2=R_{\sR_2\hbk_2}(\omega_2)R(\sR_2\hbk_2)R^{-1}(\hbk_2),\eeq
where $\bk_2$ is defined by

\beq k_2=(|\bk_2|,\bk_2)=B_z(u)\sR_1 k. \eeq
Thus we see that $B_z(u)$ appears in the decomposition of $\sR_2$
and therefore affects $\omega_2$.

Up to this point, the discussion and the formalism were purely
classical. In quantum theory, one needs the {\it unitary
representations\/} of the little group, from which those of the
complete Lorentz group can be derived. Each irreducible
representation corresponds to some species of elementary
particles. According to Eq.~(\ref{wdef}), the general
transformation law is~\cite{halpern,bog1,bog2,weinberg}

\beq U(\Lambda)|\bk,\sigma\9=\sum_{\sigma'}
D_{\sigma'\sigma}[W(\Lambda,k)]|\bq,\sigma'\9,\label{eqone} \eeq
where $D_{\sigma'\sigma}$ is a unitary representation of the
little group and

\beq  |\bk,\sigma\9\equiv|\bk\9\otimes|\sigma\9\eeq is an
appropriate basis. The helicity $\sigma={\bf J}\cdot\hbk$ of a
massless particle is Lorentz invariant, so that if we use it for
labelling basis states, then the sum in Eq.~(\ref{eqone}) consists
of a single term, and

\beq D_{\sigma'\sigma}=\exp(i\xi\sigma)\delta_{\sigma'\sigma},
 \label{dss} \eeq
where, for a Lorentz transformation which is a pure rotation,
$\xi$ is a function of $\sR$ and $k$ which is explicitly given by
Eq.~(\ref{defsR}).

It is experimentally known that some particles, like neutrinos (if
they are indeed massless), come with only one sign of helicity.
Others, like photons, may have it with both signs, but then the
phase angle $\xi$ for them is different. In general, different
values of $|\sigma|$ refer to different species of particles, such
as photons and gravitons. Within the present formalism, we cannot
offer an explanation why, for a given species of particles,
half-integral helicities appear with a definite sign. Of course,
there can be no helicity-statistics theorem, since we deal with a
single particle. These properties must follow from quantum field
theory for interacting fields.

An application of the above results is a {\it direct\/} derivation
of the Berry phase for massless particles.  Soon after the
introduction of Berry's phase \cite{berry:84} the latter was
derived for photons in the adiabatic approximation  \cite{chiao}
and then \cite{berry:87} for arbitrary changes in momentum.
Finally, derivations that are based on the analysis of connections
on Lie groups, and the Poincar\'e group in particular, were given
in \cite{bb,olk}.

Consider a sequence of rotations that eventually restores the
particle momentum to its original value. Its net effect is some
{\it active\/} rotation around the momentum's direction,
$R_{\hbk}(\omega)$. According to Eqs.~(\ref{rdeco}) and
(\ref{dss}), the helicity eigenstates acquire phases
$-\omega\sigma$, where the minus sign arises from the fact that
the transformation law (\ref{dss}) is for passive rotations.  To
relate this phase to the area on the unit sphere that is enclosed
by the orbit of $\hbk$, consider an auxiliary unit vector $\bv$ in
the plane perpendicular to $\hbk$. It is tangent to the sphere at
the endpoint of $\hbk$. Let the ortho\-normal triad $\hbk$, $\bv$,
and $\bw=\hbk\times\bv$ be parallel-transported along that orbit.
When the latter closes, the ortho\-normal triad does not return to
itself, but ends up as another triad at the same point, which is
rotated by the angle $\omega$ in the $\bv\bw$ plane. Owing to
properties of the holonomy group \cite{fra}, the rotation angle
$\omega$ is related to the spherical angle $\Omega$ by

\beq \omega=\int\! K dS=\int\! d\Omega=\Omega, \eeq
where $K$ is the Gaussian curvature, which equals 1 for the unit
sphere. The above integral is over the area enclosed by the
trajectory of $\hbk$, and the sign of $\Omega$ depends on the
orientation of the trajectory. We thus obtain

\beq \omega=\Omega. \eeq

As another application, consider the concept of reduced density
matrices \cite{qt} which is fundamental in quantum information
theory. Their properties are significantly modified by
relativistic effects
\cite{jmo,rmp}. For massless particles that admit both signs of
helicity, such as photons, a generic one-particle state is

\beq |\Psi\9=\int d\mu(k)\sum_\sigma f_\sigma(\bk)\,|\bk,\sigma\9,
\eeq where $d\mu(k)=d^3\bk/(2\pi)^3(2|\bk|)$ is a
Lorentz-invariant measure. Then the reduced density matrix for
helicity, according to the usual rules, would be

\beq \rho_{\sigma\tau}=\int d\mu(k)\,f_\sigma(\bk)f^*_\tau(\bk).
\eeq 
However, since $\xi$ in Eq.~(\ref{dss}) depends on the photon momentum
(even for ordinary rotations) the standard density matrix given by
Eq.~(\theequation) has no transformation rule at all. This makes
the standard density matrix a useless concept, even when only a
fixed reference frame is considered, since any POVM that describes
an experimental setup must have definite transformation properties
at least under ordinary rotations. It is only possible to define an
``effective'' density matrix which depends on the detection method
\cite{jmo,rmp}. This behavior contrasts with that of massive particles,
for which the little group is SO(3) and reduced density matrices behave
properly under rotations, while there is no transformation law only
under boosts \cite{pst}.

The absence of {\it any\/} Lorentz transformation law for $\rho$
is due to the fact that the momenta $k$ transform linearly, but
the law of transformation of helicity depends explicitly on $k$.
When we compute $\rho$ by summing over momenta, all knowledge of
them is lost and it is then impossible to obtain the new $\rho$ by
transforming the old one.  There is an analogous situation in
classical statistical mechanics: a Liouville function can be
defined in any Lorentz frame \cite{balescu}, but it has no
definite transformation law from one frame to another. Only the
complete dynamical system has a transformation law \cite{open}.

In summary, we have shown how apparently disparate notions ---
Wigner's little group and Berry's phase --- are closely related.
It is curious that the proof made repeated use of ordinary
rotations, namely the SO(3) group, which is by itself {\it
another\/} little group of the Lorentz group.

\bigskip  Work by AP was supported by the Gerard Swope Fund and the Fund
for Promotion of Research.

\end{document}